\documentclass[10pt]{article} 
\usepackage[T1]{fontenc}
\usepackage[utf8]{inputenc}
\usepackage[english]{babel} 
\usepackage{hyperref}

\usepackage{empheq}
\usepackage{graphicx} 
\usepackage{dsfont}
\usepackage{tikz}
\usepackage{braids}
\usepackage{graphicx}
\usepackage{wrapfig}

\usepackage{amsmath}
\usepackage{amsthm}
\usepackage{booktabs} 
\usepackage{array} 
\usepackage{paralist} 
\usepackage{verbatim} 
\usepackage{subfig} 

\usepackage{fancyhdr} 
\numberwithin{equation}{section}
\usepackage{sectsty}
\allsectionsfont{\sffamily\mdseries\upshape} 

\usepackage[nottoc,notlof,notlot]{tocbibind} 
\usepackage[titles,subfigure]{tocloft} 

\definecolor{myblue}{rgb}{.8, .8, 1}

\begin{document}

\noindent

{\bf
{\Large \textcolor{black}{Towards} a new proposal for the time delay in gravitational lensing 
}
} 

\vspace{.5cm}
\hrule

\vspace{1cm}

\noindent

{\large\bf{Nicola Alchera\footnote{\tt nicola.alchera@ge.infn.it }, Marco Bonici\footnote{\tt marco.bonici@ge.infn.it }
and Nicola Maggiore\footnote{\tt nicola.maggiore@ge.infn.it }
\\[1cm]}}

\setcounter{footnote}{0}

\noindent
{{}Dipartimento di Fisica, Universit\`a di Genova,\\
via Dodecaneso 33, I-16146, Genova, Italy\\
and\\
{} I.N.F.N. - Sezione di Genova\\
\vspace{1cm}
	\section*{Abstract }
One application of the Cosmological Gravitational Lensing in General Relativity is the measurement of the Hubble constant $H_0$ using the time delay $\Delta t$ between multiple images of lensed quasars. \textcolor{black}{This method has already been applied, obtaining a value of $H_0$ compatible with that obtained from the SNe 1A,  but non compatible with that obtained studying the anisotropies of the CMB.} This difference could be a statistical fluctuation or an indication of new physics beyond the Standard Model of Cosmology, so it desirable to improve the precision of the measurements. \textcolor{black}{At the current technological capabilities it is possible to obtain} $H_0$ to a percent level uncertainty, so a more accurate theoretical model could be necessary in order to increase the precision about the determination of $H_0$. The actual formula which relates $\Delta t$ with $H_0$ is approximated; in this paper we expose a proposal to go beyond the previous analysis and, within the context of a new  model, we obtain a more precise formula than that present in the Literature.

\vspace{2cm}\noindent Keywords: classical general relativity; gravitational lenses

\newpage

\section{Introduction}

\textcolor{black}{One of the nicest consequences of the existence of symmetries in nature is General Relativity. In fact,  the Einstein equations}
\begin{equation}
R_{\mu\nu}-\frac{1}{2}Rg_{\mu\nu}+\Lambda g_{\mu\nu}=0,
\label{einsteineq}\end{equation}
\textcolor{black}{where $R_{\mu\nu}$ and $R$ are the Ricci tensor and the Ricci scalar, respectively, $g_{\mu\nu}$ is the metric and $\Lambda$ is the cosmological constant, are the equations of motion for $g_{\mu\nu}$, seen as dynamical tensor field, naturally derived from the Hilbert action}
\begin{equation}
S_H=\int d^4x\sqrt{-g}(R-2\Lambda),
\label{hilbert}\end{equation}
\textcolor{black}{where $g$ is the determinant of $g_{\mu\nu}$. The Hilbert action \eqref{hilbert}, in turn, is the most general scalar functional, including up to second order derivatives of $g_{\mu\nu}$, invariant under diffeomorphisms of the metric $g_{\mu\nu}$
}
\begin{equation}
\delta g_{\mu\nu}={\cal L}_Vg_{\mu\nu}=\nabla_\mu V_\nu + \nabla_\nu V_\mu,
\label{diff}\end{equation}
\textcolor{black}{where $\nabla_\mu V_\nu$ is the covariant derivative of a vector field $V_\nu$ generating the diffeomorphisms. 
The transformations \eqref{diff} represent gauge transformations, whose geometrical setup is commonly exploited to obtain nontrivial results in several branch of theoretical physics, from gravity to condensed matter and AdS/CFT \cite{Blasi:2015lrg,Blasi:2017pkk,Blasi:2011pf,Blasi:2008gt,Amoretti:2013nv,Amoretti:2017xto,Amoretti:2014kba}
As it is well known, General Relativity is, under any respect, a gauge field theory, for the gauge invariance \eqref{diff}, with all the subtleties which this implies \cite{Carroll:2004st}. It is therefore perfectly legitimate to include General Relativity as a majestic consequence of the Symmetry Principle governing our Universe.}

One of the first tests of General Relativity was the effect called \textit{Gravitational Lensing} (GL): the presence of a massive object, which could be a star, a black hole or a galaxy cluster (we will refer to them as \textit{lenses}), deforms the spacetime in its neighborhood, causing the deflection of light. Although in this paper we will consider the deformation induced by massive objects, this is not the only possibility to deform the spacetime.\par
This deflection generates multiple images of the source: \textcolor{black}{according to the equations of General Relativity the photons  follow} different paths from the source to the observer.\par
The deflection of light is not the only consequence of GL because if we consider two photons, emitted at the same time but following different paths, they will be observed at different times: we will call this difference \textit{time delay}.\par
\textcolor{black}{
This delay is important because it is directly related to the value of the Hubble constant, providing us a method to determine its value. As pointed out in \cite{Efstathiou:1998xx}, there is a certain degeneracy in the determination of the cosmological parameters from the CMB \cite{Ade:2015xua} and independent measurements are important because they could break this degeneracy. In particular, the value of $H_0$ can be determined using the GL \cite{Suyu:2016qxx}\cite{Sluse:2016owq}\cite{h0licow3}\cite{Wong:2016dpo}\cite{Bonvin:2016crt} , \textcolor{black}{following \cite{Refsdal:1964nw},} or  Standard Candles \cite{Riess:2016jrr}; these measurements are compatible with each other but not with the one in \cite{Ade:2015xua}. } In order to face this problem, there have been different proposal involving, for example, dynamical dark energy \cite{DiValentino:2017iww}. 
\textcolor{black}{In order to evaluate the delay between the detection of this two photons, we should compare the \textit{flight time} needed to travel the different paths from the emitting source (S) to the observer on Earth (E). To do this, we should solve the geodesic of the photons, which in general is a tough task. We will instead adopt a perturbative approach.} 

The paper is organized as follows:
\begin{itemize}
	\item{In section \ref{standard}, in order to \textcolor{black}{face the task of solving the geodesics, the delay will be split in two contributions in order to get an approximate expression, following the standard analysis.}}
	\item{In section \ref{extension} we extend in an easy way the standard analysis.}
	\item{In section \ref{bo} we propose an alternative method to calculate the time delay, possibly in a more precise way. \textcolor{black}{This is important because, if we will obtain an expression of the delay which refines and contains the standard one, we will strengthen the result in \cite{Bonvin:2016crt}.}}
\end{itemize}

\section{Standard analysis}\label{standard}

\subsection{Basics of Gravitational Lensing}

\textcolor{black}{
We have to solve the Einstein equations \eqref{einsteineq} where the role of matter is covered by the gravitational lens L. In order to do that, we will adopt} a perturbative approach decomposing the metric $g_{\mu\nu}$ as follows
\begin{equation}
g_{\mu\nu}=\bar{g}_{\mu\nu}+h_{\mu\nu}
\end{equation}
where $\bar{g}_{\mu\nu}$ is the \textit{background metric} and $h_{\mu\nu}$ the \textit{perturbation} induced by the massive object.\par
\textcolor{black}{
In Cosmology, the commonly used energy-momentum tensor corresponding to gravitational lenses is that of non-relativistic matter, which is parametrized as a perfect fluid 
\begin{equation}
	T_{\mu\nu}=(\rho+P)U_\mu U_\nu+P g_{\mu\nu},
	\label{eqn::fluid}
\end{equation}
where the pressure $P$ is negligible with respect to the density $\rho$
\begin{equation}
P\ll\rho.
\end{equation}
hence the energy-momentum tensor in the Einstein equations for GL is}
\begin{equation}
T_{\mu\nu}=\rho U_\mu U_\nu
\label{eqn::dust}
\end{equation}
where $U_\mu$ is the  4-velocity of the lens.\par 
\textcolor{black}{The details of calculations can be found} in \cite{Carroll:2004st},  here we will simply sketch the method and expose the \textcolor{black}{main} results.\par
We are interested in the Cosmological Lensing and so we should use as background metric the Robertson-Walker (RW) metric; however we will use the Minkowski metric 
\begin{equation}
ds^2=-dt^2+dx^idx^j\delta_{ij}
\label{eqn::unperturbed}
\end{equation}
because the calculations are simpler and we will be able to insert in the result the information of the cosmological expansion. \textcolor{black}{In any case, as we will see later, the same results can be rigorously obtained perturbing the (flat) Robertson-Walker metric, as it should be. Using as background metric the Minkowski metric the result is}
\begin{equation}
ds^2=-\left( 1+2\Phi \right) dt^2+\left( 1-2\Phi\right) dx^idx^j\delta_{ij}
\label{eqn::perturbed}
\end{equation}
with $\Phi$ satisfying the Poisson equation
\begin{equation}
\nabla^2\Phi=4\pi G\rho
\label{eqn::poisson}
\end{equation}
thus we can interpret $\Phi$ as the Newtonian potential associated to the lens.\par
This result explains why we observe only two images of the source if we consider a spherically symmetric lens. In this case, the potential will be of the form
\begin{equation}
	\Phi=\Phi(r)
\end{equation}
thus the metric \eqref{eqn::perturbed} has a rotational invariance, so the angular momentum of the photon is conserved and this means that the motion of the photon is restricted to the plane individuated by the $S$, $L$ and the momentum of the photon, as in the case of the  Schwarzschild's geodesics. Furthermore, the equation which determines the position of the images, the \textit{lens equation} which can be found in \cite{Carroll:2004st}, is a quadratic equation and thus there will be two solutions.\par

\begin{figure}[h]
	\includegraphics[scale=0.34]{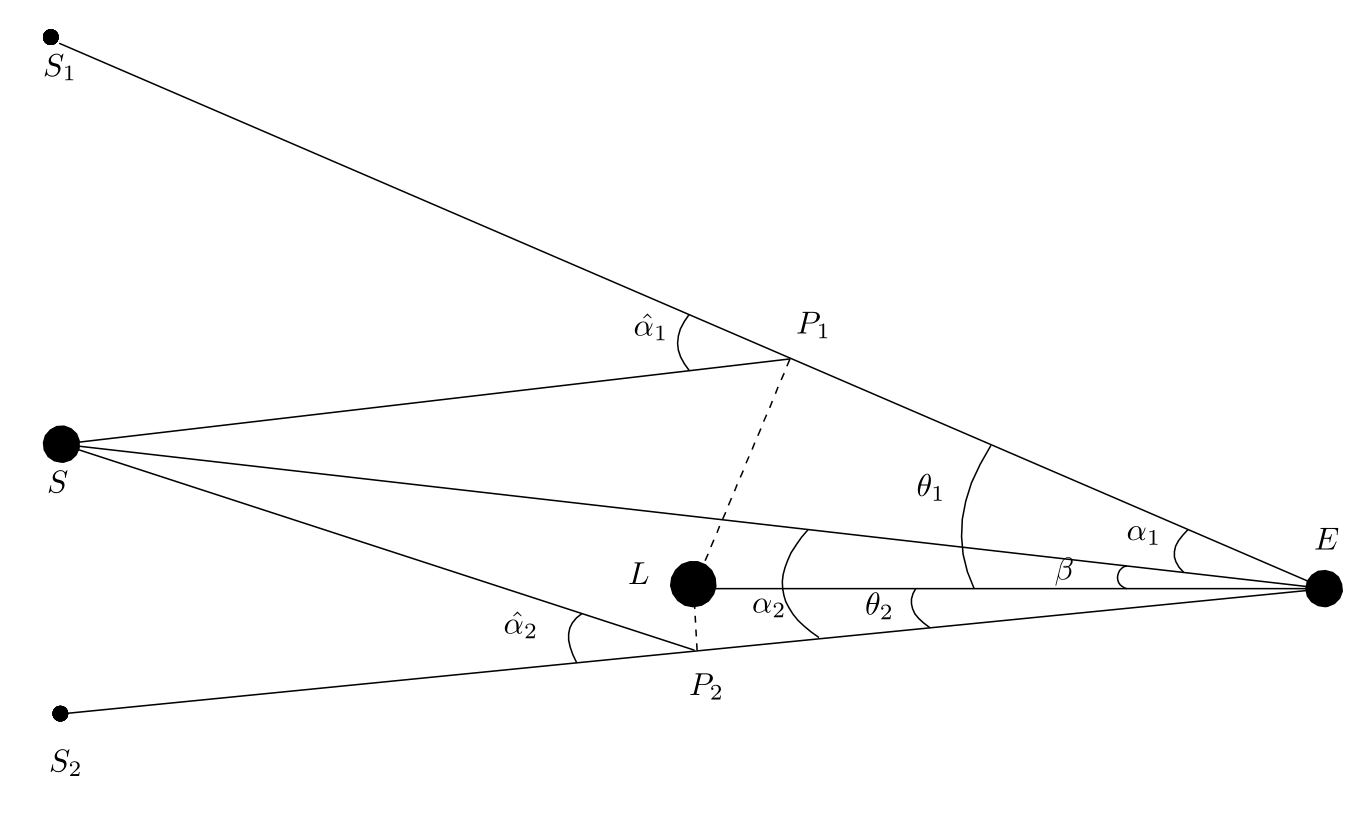}
	\caption{\textcolor{black}{$S_1$ and $S_2$ are the images of the source $S$. The points $P_1$ and $P_2$ are the deflection points of the light rays deflected by the lens $L$ and observed in $E$. $SP_1E$ and $SP_2E$ approximate the deflected photon geodesics.}}
	\label{cosmolens}
	\centering
\end{figure}
As already anticipated in the introduction, \textcolor{black}{the delay will be split} in two different parts
\begin{itemize}
	\item{\textcolor{black}{The Shapiro delay, or potential time delay, caused directly by the motion of the light through the gravitational potential of the lens}}
	\item{\textcolor{black}{The geometric delay, caused by the increased length of the total light path from the source to the earth.}}
\end{itemize}

\subsection{The Shapiro time delay \textcolor{black}{in Minkowski metric}}
\label{shapiro}

We want to study the geodesic of a photon moving in the metric \eqref{eqn::perturbed}. \textcolor{black}{Following a} perturbative approach, we will divide the geodesic in two parts, the background term $\bar{x}^{ \mu}$ and a perturbative term $x^{\prime \mu} $\footnote{From now on we will indicate with a bar all the background quantities and with a prime the perturbed quantities.}. Then we have
\begin{equation}\label{e6}
x^\mu(\lambda)=\bar  x^{\mu} (\lambda)+x^{\prime \mu} (\lambda)
\end{equation}
	where $\lambda$ \textcolor{black}{parametrizes} the geodesic. From now on we will perform all the integrals along the background paths; this is a good approximation, as long as it is satisfied
\textcolor{black}{
\begin{equation}
x^{\prime i}\partial_i\Phi\ll\Phi
\label{eqn::approx}
\end{equation}}
\textcolor{black}{This condition ensures that the potential
along the background path does not sensibly differ from that of the real path.}\par 
\textcolor{black}{The equation for null geodesic is}
\begin{equation} \label{e8}
g_{\mu \nu}  {dx^\mu  \over d\lambda} {dx^\nu \over d\lambda} = 0
\end{equation}
We will solve Eq. \eqref{e8} \textcolor{black}{perturbatively} order by order. It will be useful to define the following quantities
\begin{equation}
	k^\mu \equiv {d \bar x^\mu \over d\lambda}\qquad l^\mu \equiv {d x^{\prime \mu} \over d\lambda} 
\end{equation}
At zeroth order we have
\begin{equation}\label{e9}
\eta_{\mu \nu} {d\bar x^\mu \over d\lambda}{ d\bar x^\nu \over d\lambda} = 0
\end{equation} 
which gives us the constraint
\begin{equation}\label {eqn::vincolo}
-(k^0)^2 +  |\vec {k}|^2 =0
\end{equation} 
From now on we will use the following notation
\begin{equation}
	|\vec {k}|^2=k^2
\end{equation}
At first order we have
\begin{equation}
2\eta_{\mu \nu} k^\mu l^\nu + h_{\mu \nu} k^\mu k^\nu = 0
\end{equation}
which, using \eqref{eqn::unperturbed}, \eqref{eqn::perturbed} and \eqref{eqn::vincolo}, becomes
\begin{equation}\label{ordine1}
- k l^0 +\vec{ l} \cdot \vec{ k} = 2 k^2 \Phi 
\end{equation}
Now, let us consider the geodesic equation
\begin{equation}\label{geodetiche}
{d^2 x^\mu \over d\lambda^2} + \Gamma^\mu _{\rho \sigma} {d x^\rho \over d\lambda}{dx^\sigma \over d\lambda} =0
\end{equation}
where $\Gamma^\mu_{\rho\sigma}$ are the Christoffel symbols corresponding to the metric \eqref{eqn::perturbed}, \textcolor{black}{which can be found  in Appendix \ref{christoffel}. At order zero} we have
\begin{equation}
{d k^\mu \over d\lambda }=0
\end{equation}
This means that the background trajectories are straight lines, as we expected.\par 
At first order we have
\begin{equation}
{d l^\mu \over d\lambda}= - \Gamma^\mu _{\rho \sigma} k^\rho k^\sigma
\end{equation}
Let us consider the $\mu=0$ component
\begin{equation}\label{temporale}
{d l^0 \over d\lambda}=-2 k (\vec k \cdot \vec \nabla \Phi)
\end{equation}
and the spatial components
\begin{equation}\label{spaziale}
{d \vec{l} \over d\lambda}=-2 k^2 \nabla _{\perp} \Phi
\end{equation}

where we have introduced  the \textit{transverse gradient} $\nabla _\perp \Phi $, defined as the total gradient less the gradient along the path
\begin{equation}
\nabla_{\perp} \Phi \equiv \nabla \Phi - \nabla_\parallel \Phi  = \nabla \Phi- {1 \over k^2 } (\vec k \cdot  \nabla \Phi) \vec k
\end{equation}
It is worth emphasizing that evaluating the following indefinite integral
\begin{equation}
\begin{split}\label{l0}
l^0&= \int {dl^0 \over d\lambda } d\lambda =   - 2k  \int (\vec \nabla \Phi \cdot \vec k)    d\lambda =\\
&= -2k \int {d \vec{\bar{x}} \over d\lambda} \cdot \vec \nabla \Phi d\lambda = -2k \int \vec \nabla \Phi \cdot d \vec{\bar{x}} = -2k\Phi
\end{split}
\end{equation}
the integration constant is fixed demanding that $l_0=0$ when $\Phi=0$.
Plugging this expression in \eqref{ordine1} we obtain
\begin{equation} 
\vec l \cdot \vec k = 0
\end{equation}
which means that the two vectors are orthogonal one to each other.
\par
We can now evaluate the time delay between a photon moving \textcolor{black}{in the unperturbed Minkowski metric \eqref{eqn::unperturbed} and one moving in the perturbed metric \eqref{eqn::perturbed}}.
\textcolor{black}{
Following \cite{defalco}, let us consider a photon emitted in S, which is detected in E after being deflected by L (see Figure 1), in the perturbed metric \eqref{eqn::perturbed}. Having in mind that the approximate path travelled by the photon is SPE, where P is the deflection point closest to the lens L,} 
the flight time of the photon moving in the perturbed metric is
\begin{equation}
t = \int  {dx^0 \over d\lambda } d \lambda = \int \left(  {d \bar x^0 \over d\lambda }+{d  x^{\prime 0} \over d\lambda } \right) d \lambda= \int\left(  k^0 +l^0\right)  d\lambda 
\end{equation}
while the flight time of the photon moving in the unperturbed metric is
\begin{equation}
\bar t = \int {d \bar x^0 \over d\lambda} d\lambda = \int k^0 d\lambda
\end{equation}
The time delay between the two paths is   
\begin{equation}\label{potential}
\Delta t_1= t-\bar t = \int l^0 d\lambda
\end{equation}
Using the expression already obtained for $l^0$ given by \eqref{l0} we obtain
\begin{equation}
\Delta t_1= - 2 k \int \Phi d\lambda 
\end{equation} 
Using the infinitesimal line element $dl=kd\lambda$ we can write
\textcolor{black}{
\begin{equation}
\Delta t_1 = -2 \int_{SPE} \Phi dl
\end{equation}
We stress again that the integral is done over the path SPE \cite{defalco}. Notice that this time delay depends on the gravitational potential $\Phi$ of the lens, which therefore has the effect of reducing the effective speed of light  relative to propagation in vacuum. In presence of two images $S_1$ and $S_2$, we have to deal with two photons travelling two distinct paths, namely $SP_1E$ and $SP_2E$. Correspondingly, the total Shapiro time delay is given by \cite{defalco}
\begin{equation}
\begin{split}
\Delta t_S=\Delta t_2-\Delta t_1=-2\left( \int_{SP_2E}\Phi dl-\int_{SP_1E}\Phi dl\right)
\label{eqn::shapstep2}
\end{split}
\end{equation}
}
In order to put \eqref{eqn::shapstep2} in a more compact form we must introduce the \textit{angular diameter distance} and the \textit{gravitational lensing potential}.\par
If we observe from a point $P$ an object in $Q$ of proper length $l$, perpendicular to $PQ$ and with angular size $\theta$, then we define the angular diameter distance $d_A(PQ)$
\begin{equation}
d_{A}(PQ)=\frac{l}{\theta}
\label{eqn::angulardistance}
\end{equation}
in particular, it can be showed that in flat spacetime we have
\begin{equation}
d_A(PQ)=\frac{r_{PQ}}{1+z_Q}
\label{eqn::flat}
\end{equation}
where $r_{PQ}$ is the radial coordinate from $P$ to $Q$ in a coordinate system centered in $P$ and $z_Q$ is the redshift of $Q$ \textcolor{black}{with respect to P}; the details about the angular diameter distance can be found in \cite{Carroll:2004st}.\par 
\textcolor{black}{Moreover}, the gravitational lensing potential $\psi$ is given by
\begin{equation}
\psi(\vec{\theta})\equiv2\frac{d_A(LS)}{d_A(EL) d_A(ES)}\int\Phi(d_L\vec{\theta},l) dl
\label{eqn::lensingpot}
\end{equation}
where we inserted the dependence \textcolor{black}{on} $\vec{\theta}$ because the value of the angle determines the integration path, which is taken to be the spatial background geodesic in figure \ref{cosmolens}; it is worth emphasizing that this angles are vectors because, in general, we will not consider only planar angles but also angles in the space. Using this two quantities we can write the equation \eqref{eqn::shapstep2} as
\begin{align}
\Delta t_S&=-2\frac{d_A(LS)}{d_A(EL) d_A(ES)}\frac{d_A(EL) d_A(ES)}{d_A(LS)}\left( \int_{SP_2E}\Phi dl-\int_{SP_1E}\Phi dl\right)=\\
&=-\frac{d_A(EL) d_A(ES)}{d_A(LS)}\left( \psi(\vec{\theta}_2)-\psi(\vec{\theta}_1)\right).
\label{eqn::shaptosto}
\end{align}
\textcolor{black}{We have not yet considered the contribution arising from the expansion of the universe. However, this can be taken into account as follows.} As we can see from \eqref{eqn::shapstep2} the main contribution to the integral is originated near the lens, so we can say that the Shapiro delay is originated near the lens. This means that when photons \textcolor{black}{leave the region of space perturbed by} the lens they have already acquired the delay given by \eqref{eqn::shaptosto}, then we simply have to redshift the result by $(1+z_L)$ and we can conclude that the Shapiro time delay $\Delta t_S$ observed from the Earth is 
\begin{equation}
	\Delta t_S=-(1+z_L)\frac{d_A(EL) d_A(ES)}{d_A(LS)}\left( \psi(\vec{\theta}_2)-\psi(\vec{\theta}_1)\right)
	\label{eqn::shapiro}
\end{equation}
where we have used the definition of redshift $z$ 
\begin{equation}
a(t)=\frac{1}{1+z},
\label{eqn::redshift}
\end{equation}
\textcolor{black}{and $a(t)$ is the scale factor at time $t$. More details about redshift can be found in \cite{Carroll:2004st}. As we will see, the same result \eqref{eqn::shapiro} can be obtained perturbing the flat RW metric, with the advantage that the redshift scaling $(1+z_L)$ will be obtained naturally. and not put by hand as we just did here.}\par

\subsection{Geometric time delay}
\label{geometric}

Let us calculate the geometric time delay $\Delta t_G$. Using the lightlike interval and the unperturbed RW flat metric
\begin{equation}
ds^2=-dt^2+a^2(t) dx^idx^j\delta_{ij}
\label{eqn::unperturbedRW}
\end{equation} we have
\begin{equation}
\int_{t_S}^{t_{E_0}}\frac{dt}{a(t)}\equiv\sigma_{SE}
\label{eqn::unpgeo}
\end{equation}
where $\sigma_{SE}$ is the proper length between Earth and the light Source, $t_S$ is the emission time and $t_{E_0}$ is the arrival time of the photon running along the straight path. We perturbed the flat RW metric because it is compatible with the experimental \textcolor{black}{result $|\Omega_c|<0.1$ \cite{Carroll:2004st}}.\par 
Now, let us calculate the flight time of the photon running along the lengthened path in the perturbed metric: we can parametrize the trajectory with two segments, one from the source to the minimum distance point $P$ and one from  $P$ to the Earth (see figure \ref{cosmolens}). Thus
\begin{equation}
\int_{t_S}^{t_E}\frac{dt}{a(t)}=\sigma_{SP}+\sigma_{PE}
\label{eqn::pertgeo}
\end{equation}
We can calculate the delay $\Delta t'$ between the two paths subtracting \eqref{eqn::unpgeo} from \eqref{eqn::pertgeo}
\begin{equation}
	\int_{t_S}^{t_E}\frac{dt}{a(t)}-\int_{t_S}^{t_{E_0}}\frac{dt}{a(t)}=\sigma_{SP}+\sigma_{PE}-\sigma_{SE}
	\label{eqn::sottrazione}
\end{equation}
We can evaluate the left hand side of \eqref{eqn::sottrazione}
\begin{equation}
\int_{t_S}^{t_E}\frac{dt}{a(t)}-\int_{t_S}^{t_{E_0}}\frac{dt}{a(t)}=	\int_{t_{E_0}}^{t_E}\frac{dt}{a(t)}\approx\frac{\Delta \tilde{t}}{a(t_E)}=\Delta \tilde{t}
\end{equation}
where we used the observation that time delay is small compared to Hubble time, so we can consider $a(t)$ constant, the usual normalization $a(t_E)=1$ and we have introduced the delay between the two photons $\Delta \tilde{t}$. In order to evaluate the proper distance it is convenient to use radial coordinates with the origin positioned on the Earth, so we can immediately write
\begin{equation}
\sigma_{SE}=\int_{0}^{r_{ES}}dr=r_{ES}\qquad\sigma_{PE}=\int_{0}^{r_{EP}}dr=r_{EP}
\end{equation}
$\sigma_{SP}$ is not purely radial; from the geometry in figure \ref{cosmolens} we have
\begin{equation}
\sigma_{SP}=\sqrt{r_{ES}^2+r_{EP}^2-2r_{ES}r_{EP}\cos\alpha}
\label{eqn::cosines}
\end{equation}
We are interested in small angles, so we can perform an expansion
\begin{equation}
	\begin{split}
	\sigma_{SP}&\approx\sqrt{r_{ES}^2+r_{EP}2-2r_{ES}r_{EP}+r_{ES}r_{EP}\alpha^2}=\\
	&=(r_{ES}-r_{EP})\sqrt{1+\frac{r_{EP}r_{ES}\alpha^2}{(r_{ES}-r_{EP})^2}}=\\
	&\approx r_{ES}-r_{EP}+\frac{r_{EP}r_{ES}\alpha^2}{2(r_{ES}-r_{EP})}
	\end{split}
	\label{eqn::approximation}
\end{equation}
from which it follows
\begin{equation}
\Delta \tilde{t}=\frac{r_{ES}r_{EP}\alpha^2}{2(r_{ES}-r_{EP})}
\label{eqn::delay}
\end{equation}
We can use $r_{ES}-r_{EP}\approx r_{LS}$ because a more precise treatment would introduce higher order corrections. Thus, we have
\begin{equation}
\Delta \tilde{t}=\frac{r_{ES}r_{EP}\alpha^2}{2r_{LS}}=(1+z_L)\frac{d_A(ES)d_A(EL)\alpha^2}{2d_A(LS)}
\label{eqn::thetaquadro}
\end{equation}
where we have used \eqref{eqn::flat}.\par 
As in the previous case, we are not interested in the delay given by \eqref{eqn::thetaquadro} since it is not observable, but in the delay between two photons running along different geometric paths, so we obtain
\begin{equation}
	\Delta t_G=\Delta \tilde{t}_2-\Delta \tilde{t}_1=(1+z_L)\frac{d_A(ES)d_A(EL)}{2d_A(LS)}(\alpha_{2}^2-\alpha_{1}^2)
	\label{eqn::geom}
\end{equation}
Adding \eqref{eqn::shapiro} to \eqref{eqn::geom} we obtain the total delay $\Delta t$ 
\begin{equation}
	\Delta t=\Delta t_S+\Delta t_G=(1+z_L)\frac{d_A(ES)d_A(EL)}{d_A(LS)}\left[ \frac{(\alpha_{2}^2-\alpha_{1}^2)}{2}-\left( \psi(\vec{\theta}_2)-\psi(\vec{\theta}_1)\right) \right]
	\label{eqn::imprecise}
\end{equation}
which is the same formula that can be found in \cite{defalco}; however we want an expression \textcolor{black}{which involves} $H_0$. If we use \eqref{eqn::flat} we obtain
\begin{equation}
\begin{split}
\Delta t&=\frac{r_{ES}r_{EL}}{r_{LS}}\left[ \frac{(\alpha_{2}^2-\alpha_{1}^2)}{2}-\left( \psi(\vec{\theta}_2)-\psi(\vec{\theta}_1)\right) \right]=\\
&=\frac{r_{ES}r_{EL}}{r_{ES}-r_{EL}}\left[ \frac{(\alpha_{2}^2-\alpha_{1}^2)}{2}-\left( \psi(\vec{\theta}_2)-\psi(\vec{\theta}_1)\right) \right]
\end{split}	
\label{eqn::approdo}
\end{equation}
We will use the following relation,which can be derived using the lightlike interval and the first Friedmann equation; a complete derivation can be found in \cite{Carroll:2004st},
\begin{equation}
r_{ES}=\frac{1}{H_0}\int_{0}^{z_S}\frac{dz'}{E(z')}\equiv\frac{\mathcal{R}(z_S)}{H_0}
\label{eqn::distanza1}
\end{equation}
where
\begin{equation}
E(z)=\left[ \sum_{i}\Omega_{i0}(1+z)^{n_i}\right] ^{1/2}
\end{equation}
Notice that $\mathcal{R}(z)$ is written in terms of the cosmological parameters $\Omega_{i0}$.
If we use \eqref{eqn::distanza1}, then \eqref{eqn::approdo} becomes
\begin{equation}
		\Delta t=\frac{1}{H_0}\frac{\mathcal{R}(z_S)\mathcal{R}(z_L)}{\mathcal{R}(z_S)-\mathcal{R}(z_L)}\left[ \frac{(\alpha_{2}^2-\alpha_{1}^2)}{2}-\left( \psi(\vec{\theta}_2)-\psi(\vec{\theta}_1)\right) \right]
	\label{eqn::obs}
\end{equation}

\section{\textcolor{black}{An easy extension}}
\label{extension}

Studying delay we have obtained two different contributions: the Shapiro time delay, given by equation \eqref{eqn::shapiro}, and the geometric time delay, given by \eqref{eqn::geom}.       
\textcolor{black}{When we calculated $\Delta t_G$ we made an approximation expanding \eqref{eqn::cosines} because we neglected contributes of order $\mathcal{O}(\alpha^3)$.When we calculated $\Delta t_S$ we perturbed Minkowski rather than RW metric, so we had to add manually the redshift in order to account for the expansion of the universe. In the next subsections we will show a more precise result for $\Delta t_G$ and a more rigorous calculation for the Shapiro time delay $\Delta t_S$.}

\subsection{\textcolor{black}{The extension of $\Delta t_G$}}

Let us consider equation \eqref{eqn::cosines}
\begin{align}
\sigma_{SP}&=\sqrt{r_{ES}^2+r_{EP}^2-2r_{ES}r_{EP}\cos\alpha}
\label{eqn::expansionstep1}
\end{align} expand the RHS we obtain
\begin{equation}
\sigma_{SP}=r_{ES}-r_{EP}+\frac{r_{ES}r_{EP}}{2(r_{ES}-r_{EP})}\sum_{k=1}^{+\infty}c_k\alpha^{2k}
\label{eqn::expansion}
\end{equation}
where the first coefficients are reported in appendix \ref{coefficients}. If we repeat the analysis of section \ref{geometric} using \eqref{eqn::expansion} instead of \eqref{eqn::approximation} we obtain
\begin{equation}
\Delta t=\frac{r_{ES}r_{EP}}{2(r_{ES}-r_{EP})}\sum_{k=1}^{+\infty}c_k\alpha^{2k}
\end{equation}
$r_{ES}$ and $r_{EP}$ are not observable, but we can use  \eqref{eqn::distanza1} we have
\begin{equation}
\Delta t=\frac{\mathcal{R}(z_S)\mathcal{R}(z_{P})}{2H_0(\mathcal{R}(z_S)-\mathcal{R}(z_{P}))}\sum_{k=1}^{+\infty}c_k\alpha^{2k}
\end{equation}
Thus, the geometric time delay is
\begin{equation}
\Delta t_G=\frac{\mathcal{R}(z_S)}{2H_0}\sum_{k=1}^{+\infty}c_k\left( \frac{\mathcal{R}(z_{P_2})}{(\mathcal{R}(z_S)-\mathcal{R}(z_{P_2}))}\alpha_{2}^{2k}-\frac{\mathcal{R}(z_{P_1})}{(\mathcal{R}(z_S)-\mathcal{R}(z_{P_1}))}\alpha_{1}^{2k}\right) 
\end{equation}
The distance between $P_1$ and $L$ and between $P_2$ and $L$ are small compared to cosmological scales, thus we can make the following approximation
\begin{equation}
z_{P_2}\simeq z_{P_1}\simeq z_{L}
\end{equation}
obtaining a generalization for the geometric time delay \eqref{eqn::geom}
\begin{equation}
	\Delta t_G=\frac{\mathcal{R}(z_S)\mathcal{R}(z_{L})}{2H_0(\mathcal{R}(z_S)-\mathcal{R}(z_{L}))}\sum_{k=1}^{+\infty}c_k\left( \alpha_{2}^{2k}-\alpha_{1}^{2k}\right)
	\label{eqn::allorder}
\end{equation}
Using \eqref{eqn::allorder} instead of \eqref{eqn::geom} we obtain the following expression for the total time delay
\begin{equation}
\Delta t=\frac{1}{H_0}\frac{\mathcal{R}(z_S)\mathcal{R}(z_L)}{\mathcal{R}(z_S)-\mathcal{R}(z_L)}\left[ \sum_{k=1}^{+\infty}c_k\frac{\left( \alpha_{2}^{2k}-\alpha_{1}^{2k}\right)}{2}-\left( \psi(\vec{\theta}_2)-\psi(\vec{\theta}_1)\right) \right]
\label{eqn::moreprecise}
\end{equation}
\textcolor{black}{It is easy to check that \eqref{eqn::moreprecise} includes \eqref{eqn::imprecise}, which trivially coincides with the first term of the expansion.}\par 
\textcolor{black}{Evaluating numerically the second coefficient of the expansion in \eqref{eqn::moreprecise}, in the case of the quasar Q0957+561, it has been obtained that $c_2$ is of the order of the unity, which is good for the convergence of the series, while $\alpha$ is of the order of the arcsecond, $i.e.$ $10^{-5}$ $rad$, which is a typical value for quasars. Indeed, the second order contribution is smaller than the first one by a factor of $10^{10}$; using the lenses in the CASTLES catalogue \cite{castles} it is not possible to detect this contribution. This shows that, in order to solve the tension about $H_0$, we must follow another way.}

\subsection{\textcolor{black}{The Shapiro time delay in RW metric}}

In \ref{shapiro} we obtained the value of the Shapiro delay $\Delta t_S$ on Cosmological Scales perturbing Minkowski spacetime and adding at the result the value of the redshift of the lens. In this section we want to show a derivation of $\Delta t_S$ considering the flat RW metric \eqref{eqn::unperturbedRW} and the RW metric perturbed by a massive object.\par
\textcolor{black}{The perturbed} metric can be obtained in a similar manner to \eqref{eqn::perturbed}, following the same \textcolor{black}{steps} (more details can be found in \cite{Weinberg:2008zzc})
\begin{equation}
ds^2=-\left( 1+2\Psi(x)\right) dt^2+a^2(t)\left( 1-2\Psi(x)\right) dx^idx^j\delta_{ij}
\label{eqn::perturbedRWstep1}
\end{equation}
with $\Psi$ satisfying
\begin{equation}
\nabla^2\Psi(x)=4\pi Ga^2(t)\rho(x)
\label{eqn::Poisson}
\end{equation}
where $\rho$ is the energy density of the massive object. The energy density of the non-relativistic matter behaves as \cite{Carroll:2004st}
\begin{equation}
	\rho(x)=\rho_0(\vec{x})a(t)^{-3}
	\label{eqn::matter}
\end{equation}
It can be useful to introduce
\begin{equation}
	\Phi(x)\equiv\Psi(x)a(t)
\end{equation}
Using \eqref{eqn::Poisson} and \eqref{eqn::matter} we obtain that
\begin{equation}
	\Phi=\Phi(\vec{x})
	\label{eqn::potential}
\end{equation}
Plugging \eqref{eqn::potential} in \eqref{eqn::perturbedRWstep1} we obtain
\begin{equation}
ds^2=-\left( 1+\frac{2\Phi(\vec{x})}{a(t)}\right) dt^2+a^2(t)\left( 1-\frac{2\Phi(\vec{x})}{a(t)}\right) dx^idx^j\delta_{ij}
\label{eqn::perturbedRW}
\end{equation}
with $\Phi$ satisfying the Poisson equation \eqref{eqn::poisson}.
We perturbed the \textit{flat} RW metric because it is compatible with the observations ($|\Omega_c|<0.1$).\par Now we will calculate the delay between a photon moving in \eqref{eqn::perturbedRW} and one moving in \eqref{eqn::unperturbedRW} evaluating the integral along \textcolor{black}{the path $\gamma_1$, which is the RW deformation of the minkowskian $SP_1E$},  then we will calculate the observable delay.
Using the lightlike interval and \eqref{eqn::unperturbedRW} we have
\begin{equation}
\int_{t_S}^{t_{E_0}}\frac{dt}{a(t)}=\int_{\gamma_1} dl
\label{eqn::shapsteprw1}
\end{equation}
Instead, using the lightlike interval and the perturbed flat RW metric \eqref{eqn::perturbedRW} we have
\begin{equation}
\int_{t_S}^{t_{E}}\frac{dt}{a(t)}=\int_{\gamma_1} \sqrt{\frac{1-2\Phi a^{-1}}{1+2\Phi a^{-1}}}dl\simeq\int_{\gamma_1}\left( 1-2\frac{\Phi}{a(t)}\right) d l
\label{eqn::shapsteprw2}
\end{equation}
where in the last \textcolor{black}{step} we have performed an expansion in $\Phi/a$ because  in situation of cosmological interest it has a small value.\par 
Subtracting \eqref{eqn::shapsteprw1} from \eqref{eqn::shapsteprw2} we obtain
\begin{equation}
\int_{t_S}^{t_{E}}\frac{dt}{a(t)}-	\int_{t_S}^{t_{E_0}}\frac{dt}{a(t)}=\int_{\gamma_1}\left( 1-2\frac{\Phi}{a(t)}\right) d l-\int_{\gamma_1} dl\label{eqn::shapsteprw3}
\end{equation}
The LHS of \eqref{eqn::shapsteprw3} gives the delay between the two photons
\begin{equation}
\int_{t_S}^{t_E}\frac{dt}{a(t)}-\int_{t_S}^{t_{E_0}}\frac{dt}{a(t)}=	\int_{t_{E_0}}^{t_E}\frac{dt}{a(t)}\approx\frac{\Delta t_1}{a(t_E)}=\Delta t_1
\label{eqn::lhsshap}
\end{equation}
where  we used the observation that time delay is small compared to Hubble time, so we can consider $a(t)$ constant, and the usual normalization $a(t_E)=1$. Thus we obtain
\begin{equation}
\Delta t_1=-2\int_{\gamma_1}\frac{\Phi}{a(t)}dl
\end{equation}
The \textcolor{black}{potential} delay between two photons moving in the perturbed metric is
\begin{equation}
\Delta t_S=\Delta t_2-\Delta t_1=-2\int_{\gamma_2}\frac{\Phi}{a(t)}dl+2\int_{\gamma_1}\frac{\Phi}{a(t)}dl
\label{eqn::shapstepfin}
\end{equation}
We are not able of evaluating this integrals analytically; however we can avoid this difficulty.
Let us consider two scalar functions $f(x)$ and $g(x)$ that have the same value on a interval $\Omega$, except for a interval $\Delta x_0$ around a value $x_0$, and a scalar function $a(x)$ that is nearly constant in the interval $\Delta x_0$; then, we can make the following approximation
\begin{equation}
\int_{\Omega} a(x)\left( f(x)-g(x)\right)dx	\simeq a(x_0)\int_{\Omega} \left( f(x)-g(x)\right)dx
\label{eqn::precisino}
\end{equation}
Let us come back to \eqref{eqn::shapstepfin}: the Newtonian potential evaluated along two different paths will be sensibly different only in the neighborhood of the lens; in analogy with the previous example we can write
\begin{equation}
\Delta t_S\simeq-\frac{2}{a(t_L)}\left( \int_{\gamma_2}\Phi dl-\int_{\gamma_1}\Phi dl\right)
\label{eqn::shapprecosm}
\end{equation}
where $t_L$ is the time when the photon pass near the lens. Using  \textcolor{black}{the expression for the lensing gravitational potential} \eqref{eqn::lensingpot} and  \textcolor{black}{the redshift} \eqref{eqn::redshift}, Eq. \eqref{eqn::shapprecosm} becomes
\begin{equation}
\Delta t_S=-(1+z_L)\frac{d_A(EL) d_A(ES)}{d_A(LS)}\left( \psi(\vec{\theta}_2)-\psi(\vec{\theta}_1)\right)
\label{eqn::shapbello}
\end{equation}
which is exactly the result of \eqref{eqn::shapiro}; the main advantage of this method is that we obtained the Shapiro delay $\Delta t_S$ considering the expansion of the universe \textit{ab initio} because we have perturbed RW instead of Minkowski metric. \textcolor{black}{In other words, the scale factor $(1+z_L)$ comes naturally, without need of introducing it by hand as it has been done in \eqref{eqn::shapiro}.}

\section{\textcolor{black}{Cosmological Born-Oppenheimer approximation for time delay}}
\label{bo}
\textcolor{black}{In section \ref{extension} we calculated an extension of the geometric delay, showing that it does not solve the tension about $H_0$. This leads us to develop a different approach: we will not calculate $\Delta t_S$ and $\Delta t_G$ separately, we will calculate directly the total delay in one shot using an alternative approximation for the geodesics of the photon.}

\subsection{\textcolor{black}{The idea}}

Our idea is to divide the space into a region \textcolor{black}{where the gravitational potential originated by the lens is negligible} and another with a non vanishing gravitational potential, \textcolor{black}{in close analogy with the Born-Oppenheimer approximation in non-relativistic Quantum Mechanics}. \textcolor{black}{It is worth emphasizing that the potential, in general, does not have to possess any symmetry because in the following we will not make any assumptions about $\Phi$.} We will approximate the photon spatial geodesic  with $SQPE$, as shown in fig \ref{cosmolens1}. In particular $SQ$ and $PE$ are straight lines in the region with vanishing potential and $QP$ is a curve in the region with non vanishing potential. We will calculate the flight time of the photon moving along the curve $SQPE$ using \textcolor{black}{the unperturbed  flat RW metric} \eqref{eqn::perturbedRW} only along $QP$, while \textcolor{black}{elsewhere the perturbing effect of the lens L is taken into account by} \eqref{eqn::unperturbedRW}.
\begin{figure}[h]
	\includegraphics[scale=0.145]{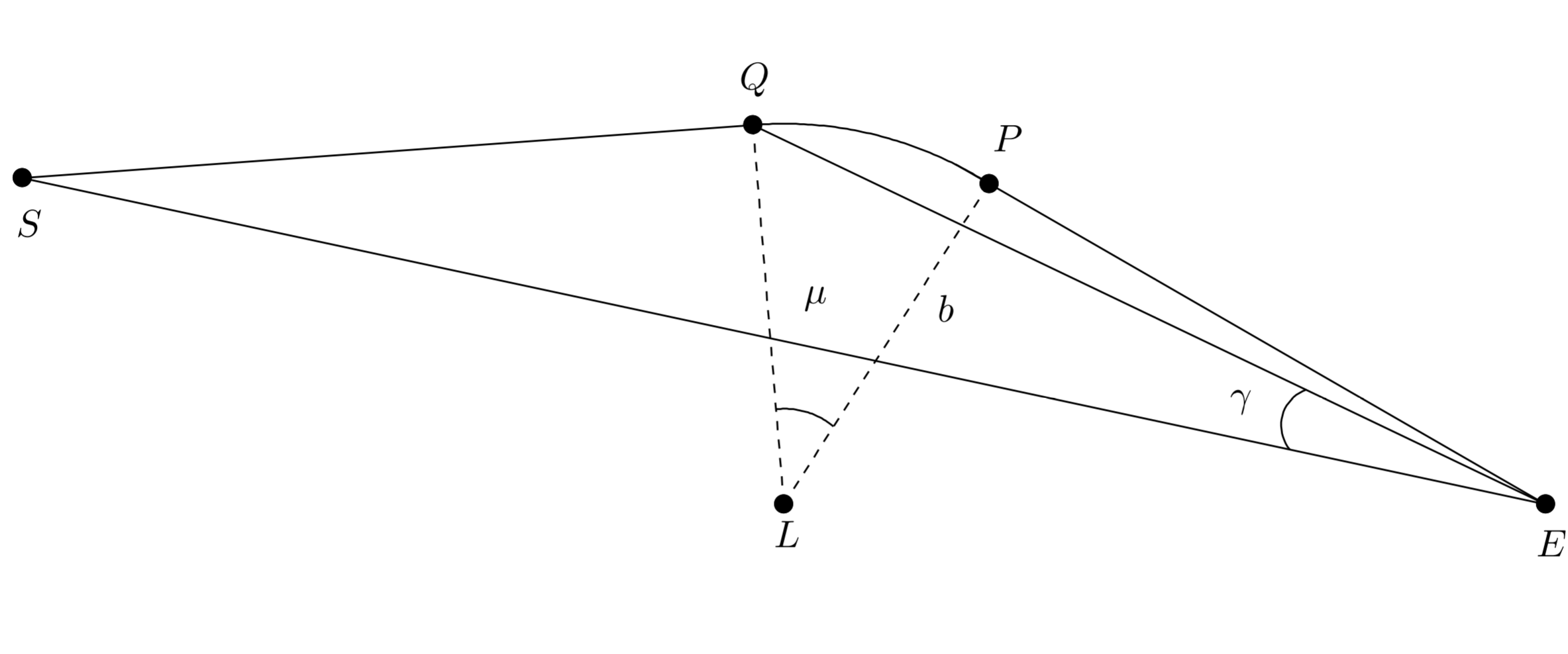}
	\centering
	\caption{The geometry we will consider.}
	\label{cosmolens1}
\end{figure}
Let us start from the photon moving in the unperturbed metric. The proper length between the Earth E and the Source S is
\begin{equation}
\int_{t_S}^{t_{E_0}}\frac{dt}{a(t)}=\sigma_{SE}
\label{eqn::dritto}
\end{equation}
Let us consider the $SQPE$ path, that we can divide \textcolor{black}{into three}  parts; using \textcolor{black}{the perturbed metric} \eqref{eqn::perturbedRW} we have
\begin{align}
&\int_{t_S}^{t_{Q}}\frac{dt}{a(t)}+\int_{t_P}^{t_{E}}\frac{dt}{a(t)}+\int_{t_Q}^{t_{P}}\frac{dt}{a(t)}=\sigma_{SQ}+\sigma_{PE}+\int_Q^P\left( 1-2\Phi a^{-1}(t)\right)dl
\label{eqn::lungopreciso}
\end{align}
Notice that the path from $Q$ to $P$ is calculated along the curved line and not along the straight line, as shown in Figure \ref{cosmolens1}.\\
Let us evaluate the left hand side of \eqref{eqn::lungopreciso};
\begin{equation}
\int_{t_S}^{t_{Q}}\frac{dt}{a(t)}+\int_{t_P}^{t_{E}}\frac{dt}{a(t)}+\int_{t_Q}^{t_{P}}\frac{dt}{a(t)}=\int_{t_S}^{t_{E}}\frac{dt}{a(t)}
\label{eqn::pippe}
\end{equation}
Instead, for the RHS of \eqref{eqn::lungopreciso}
\begin{equation}
\sigma_{SQ}+\sigma_{PE}+\int_Q^P\left( 1-2\Phi a^{-1}(t)\right)dl=\sigma_{SQ}+\sigma_{PE}+\sigma_{QP}-\frac{2}{a(t_L)}\int_{Q}^P\Phi dl
\end{equation}
So, \eqref{eqn::lungopreciso} becomes
\begin{equation}
\int_{t_S}^{t_{E}}\frac{dt}{a(t)}=\sigma_{SQ}+\sigma_{PE}+\sigma_{QP}-\frac{2}{a(t_L)}\int_{Q}^P\Phi dl
\label{eqn::pippe2}
\end{equation}
We want to calculate the time delay between the photon moving in the perturbed \textcolor{black}{RW} metric and the photon moving in the background \textcolor{black}{RW} metric; in order to obtain this result let us subtract \eqref{eqn::dritto} from \eqref{eqn::pippe2}
\begin{equation}
\int_{t_{E_0}}^{t_{E}}\frac{dt} {a(t)}=\sigma_{SQ}+\sigma_{PE}+\sigma_{PQ}-\sigma_{SE}-\frac{2}{a(t_L)}\int_{Q}^P\Phi dl
\label{eqn::approdo1}
\end{equation}
Let us evaluate the LHS of the \eqref{eqn::approdo1}: the difference between $t_E$ and $t_{E0}$ is small compared to Hubble time, thus we can consider $a(t)$ constant, and considering the usual normalization $a(t_E)=1$ we obtain
\begin{equation}
\int_{t_{E_0}}^{t_{E}}\frac{dt}{a(t)}=t_E-t_{E_0}
\label{eqn::change}
\end{equation}
We need to evaluate the RHS of \eqref{eqn::approdo1}
\begin{equation}
\sigma_{PE}=r_{PE} \qquad \sigma_{ES}=r_{ES}
\end{equation}
In order to have an explicit expression of $\sigma_{PQ}$ we can approximate it with an arc
\begin{equation}
\sigma_{PQ}=b\mu
\end{equation}
where \textcolor{black}{the angle} $\mu$ and \textcolor{black}{the distance} 
$b$ are defined in Figure \ref{cosmolens1}.
We can obtain an expression for $\sigma_{SQ}$ using the geometry in figure \eqref{cosmolens1}
\begin{equation}
\sigma_{SQ}=\sqrt{r_{EQ}^2+r_{ES}^2-2r_{ES}r_{EQ}\cos\gamma}
\end{equation}
We can use Eq. \eqref{eqn::expansion} to calculate $\sigma_{SQ}$, obtaining
\begin{equation}
\sigma_{SQ}=r_{ES}-r_{EQ}+\frac{r_{ES}r_{EQ}}{2(r_{ES}-r_{EQ})}\sum_{k=1}^{+\infty}c_k\gamma^{2k}
\end{equation}
Plugging all together we obtain
\begin{equation}
t_E-t_{E_0}=r_{ES}-r_{EQ}+\frac{r_{ES}r_{EQ}}{2(r_{ES}-r_{EQ})}\sum_{k=1}^{+\infty}c_k\gamma^{2k}+r_{EP}+b\mu-r_{ES}-\frac{2}{a(t_L)}\int_{Q}^P\Phi dl
\end{equation}
The delay between the photon moving in the perturbed metric and the photon moving in the background metric is
\begin{equation}
t_E-t_{E_0}=-r_{EQ}+\frac{r_{ES}r_{EQ}}{2(r_{ES}-r_{EQ})}\sum_{k=1}^{+\infty}c_k\gamma^{2k}+r_{Ep}+b\mu -\frac{2}{a(t_L)}\int_{Q}^P\Phi dl
\end{equation}
As in the previous cases we should consider the delay between photons running along different perturbed paths; if we define
\begin{equation}
\psi_1\left[ Q_1P_1\right] \equiv2\frac{d_A(LS)}{d_A(EL) d_A(ES)}\int_{Q_1P_1}\Phi dl
\end{equation}
and
\begin{equation}
\psi_2\left[ Q_2P_2\right] \equiv2\frac{d_A(LS)}{d_A(EL) d_A(ES)}\int_{Q_2P_2}\Phi dl
\end{equation}
we obtain
\begin{equation}
	\begin{split}
	\Delta t&=\left[ b_2\mu_2-b_1\mu_1\right] -\left[ (r_{EQ_2}-r_{EP_2})-(r_{EQ_1}-r_{EP_1})       \right]+\\
	&-(1+z_L)\frac{d_A(EL) d_A(ES)}{d_A(LS)}\left( \psi_2-\psi_1\right)+\\
	&+\left[ \frac{r_{ES}r_{EQ_2}}{2(r_{ES}-r_{EQ_2})}\sum_{k=1}^{+\infty}c_k\gamma_2^{2k}-\frac{r_{ES}r_{EQ_1}}{2(r_{ES}-r_{EQ_1})}\sum_{k=1}^{+\infty}c_k\gamma_1^{2k}\right]
	\end{split}
	\label{eqn::prototipo}
\end{equation}

using \eqref{eqn::flat} and \eqref{eqn::distanza1} we can conclude
\begin{equation}
	\begin{split}
	\Delta t&=\left[ b_2\mu_2-b_1\mu_1\right] +\frac{1}{H_0}\left[(\mathcal{R}(z_{P_2})-\mathcal{R}(z_{Q_2}))     -(\mathcal{R}(z_{P_1})-\mathcal{R}(z_{Q_1})) \right]+\\
	&+\frac{1}{H_0}\sum_{k=1}^{+\infty}\left[ \frac{\mathcal{R}(z_{S})\mathcal{R}(z_{Q_2})}{\mathcal{R}(z_{S})-\mathcal{R}(z_{Q_2})}\left( \frac{c_k\gamma_2^{2k}}{2}-\psi_2\right) -\frac{\mathcal{R}(z_{S})\mathcal{R}(z_{Q_1})}{\mathcal{R}(z_{S})-\mathcal{R}(z_{Q_1})}\left( \frac{c_k\gamma_1^{2k}}{2}-\psi_1\right) \right].
	\end{split}
	\label{eqn::alcherabonici}
\end{equation}
\textcolor{black}{The expression for the time delay} \eqref{eqn::alcherabonici} is more precise then \textcolor{black}{the one obtained in} \eqref{eqn::imprecise}. In fact, in a certain limit, the former \textcolor{black}{reduces to} the latter.
In order to see this, let us consider the following approximations
\begin{equation}
	\begin{cases}
	b_1\mu_1\simeq r_{EQ_1}-r_{EP_1}\\
	b_2\mu_2\simeq r_{EQ_2}-r_{EP_2}
	\end{cases}
	\label{eqn::approx1}
\end{equation}
\begin{equation}
	\begin{cases}
	\gamma_1\simeq\alpha_1\\
	\gamma_2\simeq\alpha_2
	\end{cases}
	\label{eqn::approx2}
\end{equation}
\begin{equation}
	z_{Q_1}\simeq z_{Q_2}\simeq z_{L}
	\label{eqn::approx3}
\end{equation}
These approximations have a precise meaning: our proposal for the time delay \eqref{eqn::alcherabonici} is more accurate than the previous one \eqref{eqn::moreprecise}, which in turn contains the ``standard'' time delay formula \eqref{eqn::imprecise}
 because we considered a more complicated geometry, but with the previous approximations we can \textcolor{black}{reduce} \eqref{eqn::alcherabonici} to \eqref{eqn::moreprecise}. \textcolor{black}{In fact,}
Plugging \eqref{eqn::approx1}, \eqref{eqn::approx2} and \eqref{eqn::approx3} in \eqref{eqn::prototipo} we find
\begin{equation}
	\Delta t=\frac{1}{H_0}\frac{\mathcal{R}(z_{S})\mathcal{R}(z_{L})}{\mathcal{R}(z_{S})-\mathcal{R}(z_{L})}\left[ \sum_{k=1}^{+\infty}\left( \frac{c_k\alpha_2^{2k}}{2}-\psi_2\right) -\sum_{k=1}^{+\infty}\left( \frac{c_k\alpha_1^{2k}}{2}-\psi_1\right) \right]
	\label{eqn::recover}
\end{equation}
There is only a small difference between  \eqref{eqn::moreprecise} and \eqref{eqn::recover}: $\psi_1$ and $\psi_2$ have not the same value of $\psi(\vec{\theta}_1)$ and $\psi(\vec{\theta}_2)$ due to the longer integration path of the latter. However, the difference is negligible because the integrand decays quickly. \textcolor{black}{Therefore, we can conclude that \eqref{eqn::alcherabonici} is an extension of \eqref{eqn::moreprecise}.}

A remark is in order concerning the points $P$ and $Q$ in figure 2: the angles in figure \ref{cosmolens} are uniquely identified unlike the angles in figure \ref{cosmolens1}. In other words, we could set the position of $Q$ and $P$ in different ways. Only after the determination of $\mu$ and $\gamma$ we will be able to use \eqref{eqn::alcherabonici}. Nevertheless, we already have some constraints: $\gamma$ must be smaller than $\theta$, while $\mu$ must be small. \textcolor{black}{However, the two points $P$ and $Q$ in figure 2 can be determined by imposing a smooth connection (for instance a tangency condition) between the straight lines  $PE$ and $SQ$ and the curve $QP$ \cite{progress}.}\par

\section{Conclusions}

In this paper we have studied one of the \textcolor{black}{main} tests of GR, the Gravitational Lensing: massive objects can modify the structure of spacetime, with the consequence that photons will not follow straight paths. This effect has a remarkable consequence: we will detect multiple images of lensed light-source, which will not be synchronized due to the different paths followed by light. In section \ref{standard} we have divided this delay in two contributions, the Shapiro, \textcolor{black}{or potential, delay} and the geometric delay, which we calculated following the standard analysis \textcolor{black}{\cite{defalco}}, obtaining an approximate expression, \eqref{eqn::imprecise}, known in the Literature \cite{defalco}. This formula is important  because it is directly related to the value of the Hubble constant $H_0$, so we can obtain a direct measurement of its value studying the time delay of lensed images. However, the results of the H0LiCOW collaboration \cite{Bonvin:2016crt}
are not compatible with the measurement obtained by the PLANCK collaboration \cite{Ade:2015xua}; this tension \textcolor{black}{is a strong motivation to improve the expression of time delay} \eqref{eqn::imprecise}. In section \ref{extension} we \textcolor{black}{studied} two slightly different approaches: we developed a more rigorous treatment for the Shapiro delay and a more precise value for the geometric delay, obtaining  \textcolor{black}{the time delay formula \eqref{eqn::moreprecise} involving higher orders in the angles $\alpha_{1,2}$, which identify the images of the source $S$. The crucial fact to notice is that it can be traced back to the Taylor series of the cosine, hence it goes like even powers of the angles. 
Now, it has been possible to give a preliminary estimate of the second order correction of the time delay formula \eqref{eqn::moreprecise}, applied to a typical source like the twin quasar Q0957+561. For this lensing phenomenon, the angular separations are of the order of one arcsecond, $i.e.$ $10^{-5}$ $rad$. Using the lens parameters, the coefficient $c_2$ in \eqref{eqn::moreprecise} is of the order of unity. 
Hence, the second order correction is of the order $10^{-10}$ which is far too small to be detected with the lenses at our disposal. For lenses with bigger angular separation (around 22 arcseconds), the second order correction reaches $10^{-8}$, which is still too little. The important conclusion is that, at least for the lenses appearing in the CASTLES catalogue \cite{castles}, the standard formula (2.54) for the time delay seems to be acceptable within the actual instrumental capabilities. This even more motivates the search for an alternative formula for time delay, which goes beyond the simple expansion in powers of the angles. }

In section \ref{bo} we \textcolor{black}{proposed a new} approach: in analogy with the first Born-Oppenheimer approximation for the scattering amplitude in non-relativistic Quantum Mechanics,  we considered the lens as a kind of cosmological scattering target, and consequently we divided the space in two regions: one where the gravitational potential originated by the lens is negligible, and another one, closer to the lens, where the gravitational potential is different from zero. This led to consider a more complicated geometry, which gave us the possibility to calculate the total delay in a single shot. \textcolor{black}{We believe that our result represent} an important improvement, because \textcolor{black}{it allows to} avoid the inaccuracies of the standard analysis. \textcolor{black}{We also checked that} the expression we have obtained for the time delay \eqref{eqn::alcherabonici} \textcolor{black}{can be reduced to, hence includes, the known result} \eqref{eqn::imprecise}.\par
In order to test the accuracy of our formula we should apply it in a real situation, obtaining an estimate of $H_0$; in particular, it would be of great interest the recognition of a situation where the difference between \eqref{eqn::imprecise} and \eqref{eqn::alcherabonici} is not negligible. \par


{\bf Acknowledgements}

It is a pleasure to thank 
{Marco Anghinolfi,
Daniele Barducci,
Gianangelo Bracco,
Lorenzo Cabona,
Roberta Cardinale,
\mbox{Anna Lucia de Marco,}
Alba Domi,
Andrea La Camera,
Davide Ricci, 
Chiara Righi, and
Silvano Tosi
}
for collaboration with us on this topic: most of what we have presented in this article has been motivated by illuminating discussions with them. In particular, we are indebted with {Gianangelo Bracco,  Alba Domi, Luca Panizzi and Silvano Tosi} for applying the formula \eqref{eqn::moreprecise} to the experimental data coming from the CASTLE database and finally again to {Luca Panizzi} for a critical and careful reading of the manuscript.
Nicola Maggiore thanks the support of INFN Scientific Initiative SFT: ``Statistical Field Theory, Low-Dimensional Systems, Integrable Models and Applications'' .
\appendix
\section{Appendix}
\subsection*{Christoffel symbols}
\label{christoffel}
The Christoffel coefficients used in \ref{shapiro} are
\begin{equation}
	\Gamma^0_{i0}=\Gamma^i_{00}=\partial_i\Phi
\end{equation}
\begin{equation}
	\Gamma^{i}_{jk}=\delta_{jk}\partial_i\Phi-\delta_{ik}\partial_j\Phi-\delta_{ij}\partial_k\Phi
\end{equation}
\subsection*{Coefficients of the expansion}
\label{coefficients}
The first coefficients appearing in the expansion present in \eqref{eqn::expansion} are
\begin{equation}
	c_1=1
\end{equation}
\begin{equation}
	c_2=-\frac{(r^2_{ES}+r_{ES}r_{EP}+r^2_{EP})}{12(r_{ES}-r_{EP})^2}
\end{equation}
\begin{equation}
	c_3=\frac{r_{ES}^4+11r_{ES}^3r_{EP}+21r_{ES}^2r_{EP}^2+11r_{ES}r_{EP}^3+r_{EP}^4}{360(r_{ES}-r_{EP})^4}
\end{equation}
\begin{equation}
	c_4=-\frac{r_{ES}^6+57r_{ES}^5r_{EP}+393r_{ES}^4r_{EP}^2+673r_{ES}^3r_{EP}^3+393r_{ES}^2r_{EP}^4+57r_{ES}r_{EP}^5+r_{EP}^6}{20160(r_{ES}-r_{EP})^6}
\end{equation}


\end{document}